\documentclass[12pt,draftcls,onecolumn]{IEEEtran}
\usepackage{graphicx}
\usepackage{epsfig}
\usepackage{algorithm}
\usepackage{algorithmic}
\usepackage{ifmtarg}
\usepackage{caption}
\usepackage{footnote}
\usepackage{cite}
\usepackage{amssymb}
\usepackage{amsthm}
\usepackage[fleqn]{amsmath}
\usepackage{amsmath}
\usepackage{epstopdf}
\usepackage{amsfonts}
\usepackage{enumitem}
\usepackage{subfigure}
\usepackage{multirow}
\usepackage{rotating}
\usepackage{hyperref}
\usepackage{color}

\usepackage[font=normalsize]{caption}
\begin{document}
\title{Improvement of the Global Connectivity using Integrated Satellite-Airborne-Terrestrial Networks with Resource Optimization}
\author{\IEEEauthorblockN{Ahmad Alsharoa, \textit{Senior Member, IEEE}, and Mohamed-Slim Alouini, \textit{Fellow, IEEE}}
\thanks { \vspace{-0.5cm}\hrule
\vspace{0.1cm} \indent Ahmad Alsharoa is with the Electrical and Computer Engineering Department, Missouri University of Science and Technology, Rolla, Missouri 65409, USA, E-mail: aalsharoa@mst.edu.
\newline \indent Mohamed-Slim Alouini is with the Computer, Electrical and Mathematical Sciences and Engineering (CEMSE) Division, King Abdullah University of Science and Technology (KAUST), Thuwal, Makkah Province, Saudi Arabia. E-mails: slim.alouini@kaust.edu.sa.
}\vspace{-.1cm}}

\maketitle
\thispagestyle{empty}
\pagestyle{empty}

\begin{abstract}
\boldmath{
In this paper, we propose a novel wireless scheme that integrates satellite, airborne, and terrestrial networks aiming to support ground users. More specifically, we study the enhancement of the achievable users' throughput assisted with terrestrial base stations, high altitude platforms (HAPs), and satellite station.
% goal is to optimize the front-hauling and back-hauling associations, transmit powers of the base stations, and the HAPs' locations in order to maximize the users' throughput utility.
The goal is to optimize the resource allocations and the HAPs' locations in order to maximize the users' throughput.
In this context, we propose to solve the optimization problem in two stages; first a short-term stage and then a long-term stage.
In the short-term stage, we start by proposing a near optimal solution and low complexity solution to solve the associations and power allocations.
In the first solution, we formulate and solve a binary linear optimization problem to find the best associations and then using Taylor expansion approximation to optimally determine the power allocations. While in the second solution, we propose a low complexity approach based on frequency partitioning technique to solve the associations and power allocations.
One the other hand, in the long-term stage, we optimize the locations of the HAPs by proposing an efficient algorithm based on a recursive shrink-and-realign process. Finally, selected numerical results show the advantages provided by our proposed optimization scheme.
}
\end{abstract}


\section{Introduction}\label{Introduction}
\subsection{Background}
\IEEEPARstart{W}ith the rapid growth in mobile and wireless devices in addition to huge data traffic, the traditional terrestrial network is expected to face difficulties in supporting the demands of the users. In addition, it is infeasible to develop terrestrial infrastructure in remote areas to provide telecommunication services. To mitigate these limitations, integrating the current terrestrial network with high altitude stations has become necessary to provide global connectivity. %aiming to help in many of today's challenges in the area of wireless communications

Integrating the terrestrial network with the space network, including satellite stations, is one of the proposed solutions in order to increase the network's coverage and capacity. Several techniques have been studied in the literature including using multiple spot beams at the satellite station associated with multiple protocol label switching and spectrum access controlled~\cite{sat1,sat2}. For instance, In~\cite{sat1}, a cooperative scheme base on non-orthogonal multiple access has been proposed, where the satellite station communicates with the users via the help of terrestrial base station (TBS) working as a relay. While in~\cite{sat2}, the authors propose to use TBSs as amplify relays to maintain the communication links between multiple satellite stations and multiple users. Furthermore, several spectrum resources techniques have been investigated recently to improve the communication link by utilizing the spectrum in efficient ways such as allocating the system resources to hot beams in order to meet user demands or to address adverse channels~\cite{sat4,sat5,sat3}.
For example, in~\cite{sat3}, the spectrum management scheme of satellite-terrestrial integration is proposed to address the unpredictable demands of users associated with the satellite station. The authors in this work introduced a hunger marketing strategy that manages the spectrum to influence users’ behavior in bandwidth planning and benefit satellite system meanwhile. Therefore, allowing a more efficient dynamic spectrum management.

However, all traditional satellite communications techniques are constrained due to the high path-loss attenuation of the communication links between the satellite stations and the ground users. In addition to that, the satellite stations can be located in various orbital heights and, thus, causing extra delay when providing real-time services to ground users.
To fill the gap, High altitude platforms (HAPs) such as airships and balloons operating in the stratosphere, at altitude of 17 Km to 20 Km, can be a promising wireless solution that is expected to work along with the existed satellite and terrestrial networks~\cite{Airborne_survey}. This altitude range is chosen because of its low wind currents and low turbulence which reduces the energy consumptions aiming to maintain the position of the HAP. Therefore, HAPs can act as aerial base stations to improve the communication links between satellite stations and ground users and hence improve the overall network throughput. This satellite-airborne-terrestrial integrated system can be part of civilian life with high-reward opportunities in improving the wireless utilization and provisioning better wireless access for public safety and first responders.

\subsection{Related Works}
HAPs acting as wireless BSs have been proposed to help the global connectivity~\cite{HAP1,HAP2}. The main advantages of using HAPs over the TBSs can be summarized as follows~\cite{Airborne_survey,integration2}:
\begin{itemize}
  \item High coverage area: The broadband coverage area of TBSs (coverage area radius around 1000 m) is usually limited compare to HAPs (coverage area radius around 30 Km) due to large non-line-of-sight pathloss. Thus, only few HAPs can cover some small countries~\cite{HAP_japan}.
      % , for example a country like Japan can be covered by approximately 16 HAPs with a minimum elevation angle of $10^o$.
  \item Dynamic and quick deployment: The HAPs have the ability to fly to infrastructure-less regions in order to enable on-demand services.
  \item Low energy consumptions: The HAPs can be equipped with solar panels that collecting energy during the daytime. Thus, with a careful trajectory optimization, HAPs can be self powered~\cite{fb_HAP,fb2_HAP}.
\end{itemize}
On the other hand, the main advantages using HAPs over the satellite stations can be summarized as follows~\cite{HAP1}
\begin{itemize}
  \item Quick and low cost deployment: HAPs can accommodate temporal and traffic demand quickly, where one HAP is enough to start the service. Also, HAPs can play a significant role in emergency or disaster relief applications by flying to desired areas, in short timely manner, in order to restart the communications~\cite{HAP_quick}. In addition, the deployment cost of the HAPs is much less than the satellite deployment cost.
  \item Low propagation delays and strong signals: Due to the high path-loss attenuation between satellite stations and ground users, HAPs can provide services to ground users in lower delay~\cite{HAP2}.
\end{itemize}

Several papers in the literature have studied the deployment of the HAPs~\cite{HAP3,fb_HAP,fb2_HAP}. For instance, the work proposed in~\cite{HAP3} investigates the HAPs' deployment taking into consideration the quality-of-service (QoS) of ground users. The authors propose a self-organized game theory model, where the HAPs are modeled as  rational and self-organized players with the goal of achieving optimal configuration of the HAPs that maximizing the users' QoS. While the work in~\cite{fb_HAP}, propose trajectory optimization techniques for HAPs equipped with solar power panels. More specifically, the authors in~\cite{fb_HAP}, proposed a greedy heuristic and realtime solution to optimize the HAPs' trajectory by minimizing the consumed energy which is constrained by the amount of the harvested energy. In~\cite{fb2_HAP}, the authors extended their work presented in~\cite{fb_HAP}, to include several trajectory optimization methods aiming to maximize the storage energy in HAPs instead of minimizing the consumed energy.

Furthermore, improving the system throughput has been considered as another important key factors in HAPs communications. The authors in~\cite{HAP_channel2}, propose multicast system model that uses orthogonal frequency division multiple access (OFDMA) to find the best transmit powers, time slots, and sub-channels in order to maximize the total ground users' throughput. In other words, the goal was to maximize the number of users that receive the requested multicast streams in the HAP service area in a given OFDMA frame. The improvement of the multiple HAPs' capacity is presented in~\cite{HAP5}, where the authors show that the HAPs can offer a spectrally efficiency by exploiting the directionality of the user antenna. In other words, the authors explained how multiple HAPs can share the same frequency band by taking the advantage of the users' antenna directionality. In~\cite{HAP6}, the authors investigated the multi-user multiple-input multiple-output (MU-MIMO) in HAP communication, where the HAPs are equipped with large-scale antenna arrays. The goal was to formulate and solve a low computational complexity technique that maximizes the signal-to-interference-noise ratio (SINR) and limits the interference between users.
However, all these works have not considered the back-hauling (BH) link for the HAPs.

Few research in the literature have considered the integration between satellite, airborne, and terrestrial networks~\cite{integration1,integration3}. For instance, the work in~\cite{integration1}, consider using HAP as a relay to enhance the satellite link. The goal was to reduce the total energy consumption while achieving certain downlink users' QoS from the satellite.
While the authors in~\cite{integration3}, present the idea of bidirectional function offloading and its possible applications by exploring the full advantages of integrated networks. More specifically, the work studied the virtual network functionality and service function chaining that enabling network reconfiguration framework.

\subsection{Contributions}
This paper studies the resource allocation with front-hauling (FH) and back-hauling (BH) associations in order to improve the global connectivity using satellite, airborne, and terrestrial networks integration. Satellite station and HAPs can play significant role in global connectivity in the case when the TBSs are overloaded or to support users with high throughput located outside the TBSs coverage areas (e.g., suburban and remote areas).
The objective of the proposed framework is to improve the downlink throughput of the ground users while respecting the resource limitations.
To the best of our knowledge, the problem of resource, power management, and associations (including FH and BH associations) for satellite, airborne, and terrestrial networks integration has not been discussed so far. The main contributions of this work can be summarized as follows:
\begin{itemize}
  \item The objective is to support the terrestrial network with satellite station and HAPs in order to enhance the network performance. In our framework, an optimization problem is formulated aiming to maximize the downlink users' throughput while taking into account into account the power budgets limitation, associations, and BH constraints.
  \item We formulate two problems based on how often optimizing the parameters: short-term stage and long-term stage. In the short-term stage, we solve the FH and BH associations and power allocations with fixed HAPs' location. In long-term stage, and based on average users distribution, we optimize the locations of the HAPs. In other words, the short-term stage variables can be optimized frequently, while the long term stage variables can be optimized in long time periods.
   \begin{itemize}
  \item Short-term stage: Due to the non-convexity of the formulated problem, we propose two solutions. In the first solution, we start by jointly optimize the FH and BH association given fixed transmit power levels. Then, optimizing the transmit powers using Taylor expansion approximations. In the second solution, we propose a low complexity solution based on frequency partitioning (FP) technique to optimize the associations and transmit powers.
  \item Long-term stage: We propose an efficient and low complexity heuristic algorithm based on shrink-and-realign (SR) process to optimize the locations of the HAPs. Note that, this algorithm will be optimized once every long time frame.
      \end{itemize}
  \item We consider different users objective functions utilities depending on the level of fairness among users.
  \item Finally, we analyze the performance of our proposed scheme under different system parameters.
\end{itemize}

\subsection{Paper Organization}
The remainder of this paper is organized as follows. Section~\ref{SystemModel} provides the system model. The problem formulation is given in Section~\ref{ProbelmFormulation}. Section~\ref{solutions} describes the short-term stage by optimizing the associations and power allocation solutions. Section~\ref{solutionsHAP} gives the long-term stage by optimizing the locations of the HAPs. The numerical results are discussed in Section~\ref{Simulation}. Finally, the paper is concluded in Section~\ref{Conclusion}.

\section{System Model}\label{SystemModel}
We consider an integrated communication network consisting of three tiers: (i) space tier with one satellite station, (ii) air tier with $L$ HAPs, and (iii) ground tier with $M$ TBS. In addition, we also consider $W$ gate ways (GWs) feeder stations to be used for the BH. We aim to provide downlink communications to $U$ ground users in a certain geographical area as shown in Fig.~\ref{SystemModel}.
In Table~\ref{Tab3}, we summarize the main notations used in this paper.
\begin{table}[h!]
\begin{center}
\caption{List of Notations}
\label{Tab3}
\begin{tabular}{|c|c|}%
\hline
   \textbf{Notation} & \textbf{Description}\\
\hline
$U$ & 	Number of users  \\
\hline
$M$ & 	Number of TBSs  \\
\hline
$L$ & 	Number of HAPs  \\
\hline
$W$ & 	Number of GWs  \\
\hline
$N^S$ & Number of available RBs at tier $S$  \\
\hline
$\bold{J}^U_u$ & User geographical coordinates  \\
\hline
$\bold{J}^S_s$  & BS $s$ geographical coordinates in tier $S$\\
\hline
$h^S_{su,n}$  & FH channel gain between BS $s$ and user $u$ over RB $n$ \\
\hline
$g_{xy}$  & BH channel gain between $x$ and $y$\\
\hline
 $A^S_{su}$ & Attenuation gain due to environment effect \\
 \hline
$\Omega_0,\Omega_1, \Omega_2$  & Shadowed Rician fading parameters \\
\hline
 $\kappa_L$  & Rician fading parameter \\
\hline
$\epsilon^S_{su,n}$  & A binary variable indicates the FH association between \\& BS $s$ and user $u$ over RB $n$\\
\hline
$\delta_{wl}$  & A binary variable indicates the BH association between \\& GW $w$ and HAP $l$\\
\hline
$P^S_{su,n}$  & FH transmit power of BS $s$ over RB $n$\\
\hline
$P_0$  & BH transmit power\\
\hline
$B^S$  & FH bandwidth of BS $s$ in tier $S$ \\
\hline
$B_0$  & BH bandwidth \\
\hline
$\mathcal{N}_0$  & Noise power \\
\hline
$R^S_{su,n}$ & FH data rate from BS $s$ to user $u$ over RB $n$ \\
\hline
$\bar{R}_{wl}$ & BH data rate from GW $w$ to HAP $l$\\
\hline
 $T$ &  Maximum  shrink-and-realign iterations\\
  \hline
  \end{tabular}
\end{center}
\end{table}

\begin{figure}[t!]
\center
\includegraphics[width=3.5in]{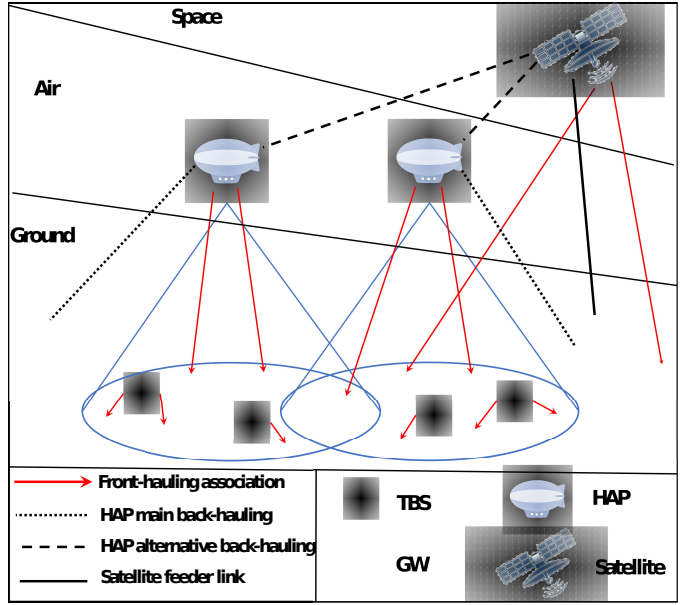}
\caption{System model.}\label{SystemModel}
\end{figure}

%\begin{figure*}
%\vspace{-2mm}
%\begin{tabular}{l}
%\begin{minipage}[t!]{0.9775\textwidth}
%\begin{center}
%\includegraphics[width=4in]{Figures/HAP_systemModel.pdf}
%\caption{System Model.}\label{SystemModel}
%\end{center}
%\vspace{1pt}
%\end{minipage}\\
%\end{tabular}
%\vspace{-15pt}
%\end{figure*}

\subsection{Assumptions}
\begin{figure}[t!]
  \centerline{\includegraphics[width=3in]{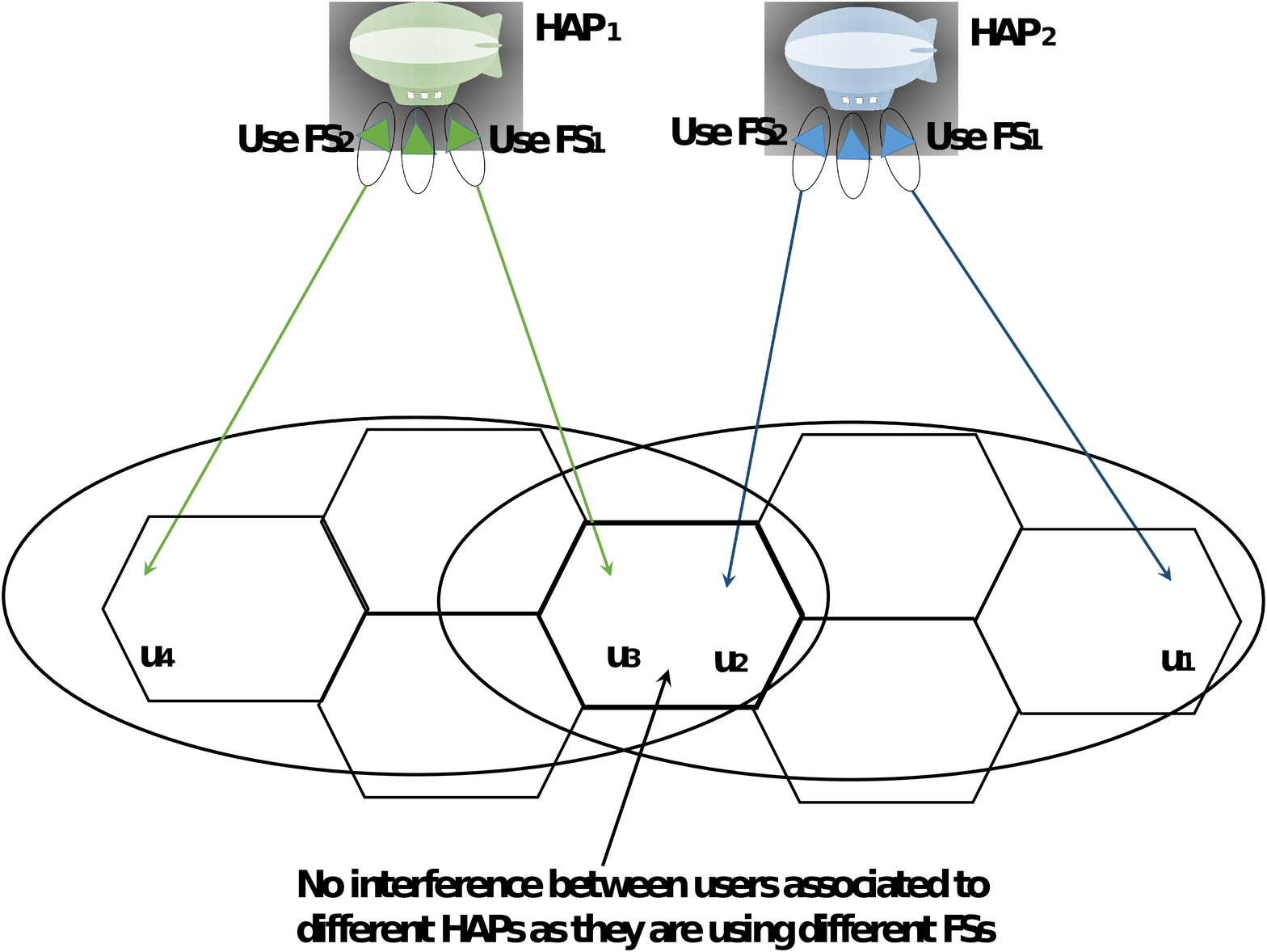}}
   \caption{Location of the HAPs.}\label{angle_b}
\end{figure}

We assume that the OFDMA technique is adopted.
The available spectrum is divided into number of resource blocks (RBs), where each RB in tier $S$ has a bandwidth of $B^S$ Hz~\cite{3GPP_release11a}. We assume that there is no interference between different tiers, where the available bandwidth is divided sparsely between these tiers.
We denote by $N^S$ the number of available RBs at tier $S$ (i.e., $S=0$ for the space tier and $S=L$ for the air tier, and $S = M$ for the ground tier).
In addition to the above assumption, we make the following practical assumptions: (i) a user is served at most by one BS\footnote{We refer to BS as TBS, HAP, or satellite station.} using only one RB. (ii) there is no intra-cell interference on the downlink between users associated to the same BS as they are using different sets of orthogonal RBs.
(iii)  an inter-cell interference between users associate to different TBSs in the ground tier (i.e., a TBS with nearby TBSs). On the other hand, we ignore this interference in the space and air tiers because the satellite station and HAPs can be equipped with different antennas and different beams where they allow managing the resources in an efficient way by using different frequency sets (FSs)~\cite{HAP5}. Note that the antennas of HAPs are arranged to produce a regular hexagonal structure, where multiple beam antenna payload at each HAP serve multiple ground cells as shown in Fig.~\ref{angle_b}.

\subsection{Channel Model}
We consider a 3D coordinate system where the coordinate of user $u$, BS $s$ are given, respectively, as $\bold{J}^U_u= [x^U_u, y^U_u, 0]'$, $\bold{J}^S_s= [x^S_s, y^S_s, z^S_s]'$, $S\in\{M:\text{for TBSs},L: \text{for HAPs},0: \text{for satellite station}\}$, where $[.]'$ denotes the transpose operator.
%
%
%TBS $m$, and UAV $l$ are given, respectively, as $\bold{J}^U_u= [x^U_u, y^U_u, 0]^T$, $\bold{J}^M_m= [x^M_m, y^M_m, z^M_m]^T$, and $\bold{J}^L_l= [x^L_l, y^L_l, z^L_l]^T$ where $[.]^T$ denotes the transpose operator. %For simplicity, let us define $S$ as the total number of TBSs or HAPs, such that $S \in \{M,L\}$.

\subsubsection{Front-hauling Channel Model}
Taking into account the fading and shadowing fluctuations in addition to the pathloss, the channel gain between BS $s$ and user $u$ can be modeled as:
\begin{equation}
\label{channelGain}
   h^S_{su,n} = \left(\frac{C}{4\pi d^S_{su} f^S_c}\right)^2 A^S_{su} F^S_{su,n},
\end{equation}
where $C$ and $f^S_c$ are the speed of light and carrier frequency in tier $S$, respectively. $d^S_{su}$ is the distance between BS $s$ in tier $S$ and user~$u$. In~\eqref{channelGain}, the first factor $\left(C/4\pi d^S_{su} f^S_c \right)^2$ captures propagation and path-loss and the second
$A^S_{su}$ is the attenuation gain due to environment effects which depends on the distance between the BS $s$ and user $u$ and given as~\cite{Affective_gain,HAP_channel2}:
\begin{equation}
  A^S_{su}=\hspace{-.1cm}\left\{
   \begin{array}{ll}
   \hspace{-.2cm}A^M_{mu}=1 \quad \quad, \hspace{-.3cm}& \hbox{for ground tier,} \\
   \hspace{-.2cm}A^L_{lu}=10^\frac{3 d^L_{lu} \chi }{10 z^L_l}, \hspace{-.3cm}& \hbox{for air tier,} \\
   \hspace{-.2cm}A^0_{0u}=10^\frac{3 d_{0u} \chi }{10 z_0}, \hspace{-.3cm}& \hbox{for space tier,}
                   \end{array}
                 \right.
\end{equation}
where $\chi$ is the attenuation factor due to environment effects such as rain, wind, cloud, etc.
The last factor in~\eqref{channelGain}, $F^S_{su,n}$, corresponds to fading distribution between BS $s$ in tier $S$ and user $u$ over RB $n$.
For the ground tier ( i.e., $S=M$, $s=m$), $F^S_{su,n}$ corresponds to Rayleigh fading power between TBS $m$ and user $u$ with a Rayleigh parameter $a$ such that $E\{|a|^{2}\} = 1$. %It also contains the the factor, $\xi_{mu}$, that captures log-normal shadowing with zero-mean and a standard deviation $\sigma_{\xi}$~\cite{Qualcomm_D2Dchannels_3GPPmeetings_MaltaFeb2013}.
For the air tier (i.e., $S=L$, $s=l$), $F^S_{su,n}$ is the Rician small scale gain between HAP $l$ and user $u$ over RB $n$ with Rician factor equal to $\kappa_L$~\cite{HAP_channel1,HAP_channel2}. Finally, for the space tier (i.e., $S=0$, $s=0$), the $F^S_{su,n}$  ia assumed as shadowed Rician fading based on ($\Omega_0,\Omega_1, \Omega_2$), where $\Omega_0$, and $2 \Omega_1$ are the average power of the line-of-sight (LoS) component and of the multipath component, respectively. $\Omega_2$ is a Nakagami parameter ranging from $0$ to $\infty$ that shows the intensity of the fading~\cite{Sat_channel}.
Without loss in generality, fast fading is assumed to be approximately constant over the subcarriers of a given RB, and independent identically distributed (i.i.d) over RBs.

\subsubsection{Back-hauling Channel Model}
We assume that the BH channel gain between $x$ and $y$ stations, i.e., between GW ($x=w$) HAP $l$ ($y=l$) for the main BH link,
and between satellite station ($x=0$) and HAP $l$ ($y=l$) for the alternative BH link, can be given as~\cite{HAP_channel1,HAP_channel2,Sat_channel2,Sat_HAP}
\begin{equation}\label{channel_Satellite}
g_{xy} = \left(\frac{C}{4\pi d_{xy} f_c}\right)^2 A_{xy} F_{xy},
\end{equation}
where $F_{xy}$, corresponds to Rician fading distribution between $x$ and user $y$ with Rician factor equal to $\kappa_{xy}$. The choice of $\kappa_{xy}$ will be depends on the main or alternative BH links.

\subsection{Front-hauling Association}
%A ground user can be associated with only one BS (i.e., either a TBS, HAP, or satellite station).
%Therefore, two binary variables are introduced. The first binary variable $\epsilon_{su,n}$, where $s \in\{m,l\}$, is equal to 1 if TBS $m$ is associated with user $u$ over the RB $n$ and 0 if HAP$l$ is associated with user $u$ over the RB $n$, and is given as follows:
%\begin{equation}
%  \epsilon^S_{su,n}=\hspace{-.1cm}\left\{
%   \begin{array}{ll}
%   \hspace{-.2cm}1, \hspace{-.3cm}& \hbox{if TBS $m$ is associated with user $u$ over RB $n$.} \\
%   \hspace{-.2cm}0, \hspace{-.3cm}& \hbox{if HAP $l$ is associated with user $u$ over RB $n$.}
%                   \end{array}
%                 \right.
%\end{equation}
%
%The second binary variable $\epsilon^0_{0u,n}$ is introduced, where it is equal to 1 if the satellite is associated with user $u$ over RB $n$ and 0 otherwise, and is given as follows:
%\begin{equation}
%  \epsilon_{0u,n}=\hspace{-.1cm}\left\{
%   \begin{array}{ll}
%   \hspace{-.2cm}1, \hspace{-.3cm}& \hbox{if the satellite is associated with user $u$ over RB $n$.} \\
%   \hspace{-.2cm}0, \hspace{-.3cm}& \hbox{other wise.}
%                   \end{array}
%                 \right.
%\end{equation}
A ground user can be associated with only one BS (i.e., either a TBS, HAP, or satellite station) at a certain time.
Therefore, we introduce a binary variable $\epsilon^S_{su,n}$, where $S \in\{M,L,0\}$, $s\in\{m,l,0\}$ for TBS $m$, HAP $l$, and satellite station, respectively, is equal to 1 if BS $s$ is associated with user $u$ over the RB $n$ and 0 otherwise, and is given as follows:
\begin{equation}
  \epsilon^S_{su,n}=\hspace{-.1cm}\left\{
   \begin{array}{ll}
   \hspace{-.2cm}1, \hspace{-.3cm}& \hbox{if BS $s$ is associated with user $u$ over RB $n$.} \\
   \hspace{-.2cm}0, \hspace{-.3cm}& \hbox{otherwise.}
                   \end{array}
                 \right.
\end{equation}

%
%
%two binary variables are introduced. The first binary variable $\epsilon^S_{su,n}$, where $S \in\{M,L\}$, $s=m$ for TBS $m$, and $s=l$ for HAP $l$, is equal to 1 if BS $s$ is associated with user $u$ over the RB $n$ and 0 otherwise, and is given as follows:
%\begin{equation}
%  \epsilon^S_{su,n}=\hspace{-.1cm}\left\{
%   \begin{array}{ll}
%   \hspace{-.2cm}1, \hspace{-.3cm}& \hbox{if BS $s$ is associated with user $u$ over RB $n$.} \\
%   \hspace{-.2cm}0, \hspace{-.3cm}& \hbox{other wise.}
%                   \end{array}
%                 \right.
%\end{equation}
%
%The second binary variable is $\epsilon_{0u,n}$, where it is equal to 1 if the satellite is associated with user $u$ over RB $n$ and 0 otherwise, and given as follows:
%\begin{equation}
%  \epsilon_{0u,n}=\hspace{-.1cm}\left\{
%   \begin{array}{ll}
%   \hspace{-.2cm}1, \hspace{-.3cm}& \hbox{if the satellite is associated with user $u$ over RB $n$.} \\
%   \hspace{-.2cm}0, \hspace{-.3cm}& \hbox{other wise.}
%                   \end{array}
%                 \right.
%\end{equation}
We assume that each BS can be associated with multiple users. On the other hand, each user can be associated with one BS at most using one RB. Hence, the following condition should be respected:
\begin{equation}
\sum_{S \in\{M,L,0\}} \sum_{s=1}^{S} \sum_{n=1}^{N^S}\epsilon^S_{su,n} \leq 1, \quad \forall u.
\end{equation}

%In each BS cell, a single user can be allocated to a one RB at most at a given transmission time interval. Hence, we have:
%\begin{equation}
%\sum_{n=1}^{N^S}\epsilon^S_{su,n} + \sum_{n=1}^{N^L}\epsilon^L_{lu,n} + \sum_{n=1}^{N}\epsilon_{0u,n} \leq 1, \quad \forall m, \forall l, \forall u.
%\end{equation}

In addition, in order to mitigate the intra-cell interference inside each cell corresponding to BS $s$ in tier $S$, we assume that different RBs are allocated to different users that associated with same BS $s$. Therefore, the following condition should be satisfied, respectively:
\begin{equation}
\sum_{u=1}^{U}\epsilon^S_{su,n} \leq 1, \quad \forall s, \forall S, \forall n=1,..,N^S.
\end{equation}

%Finally, we assume that each BS $s$ can be associated with maaximum number of users at the same time due to the limitation of its RBs. Therefore, the following constraints should be respected:
%\begin{equation}
%\sum_{n=1}^{N^S} \sum_{u=1}^{U}\epsilon^S_{su,n} \leq \bar{N}^S, \quad \forall s, \forall S.
%\end{equation}

\subsection{Back-hauling Association}
It is assumed that HAP $l$ cab be associated with either GW $w$ as a main BH link or with the satellite station as an alternative BH link based on the users distributions and the quality of the BH link as shown in Fig.~\ref{SystemModel}.
Therefore, we introduce another binary variable $\delta_{wl}$ for theBH association where $\delta_{wl}$ ($w=0$ for satellite station and $w=1,..,W$ for GWs) is equal to 1 if HAP $l$ is associated with station $w$ and 0 otherwise.
Without loss of generality, we assume that each HAP should be strictly associated with one station (either one GW or satellite station). Therefore, the following equality should be respected:
\begin{equation}
 \sum\limits_{w=0}^W \delta_{wl} = 1, \quad \forall l.
\end{equation}

Finally, we denote by $\bar{L}_w$ the maximum number of HAPs that can be associated to station $w$ in the BH, such that:
\begin{equation}
 \sum\limits_{l=1}^L \delta_{wl} \leq  \bar{L}_w \quad \forall w.
\end{equation}

\subsection{Downlink Data Rates}\label{Downlink_Data_rate}
It is assumed that the total spectrum is shared sparsely between tiers. The achievable FH data rate of user $u$ served from a BS $S$, where $S \in\{M,L,0\}$, over RB $n$ can be given as:
\begin{equation}\label{rate}
  R^S_{su,n}=B^S \epsilon^S_{su,n}  \log_2\left(1+\frac{P^S_{su,n}\,h^S_{su,n}}{\mathcal{I}^S_{su}+\mathcal{N}_0 B^S}\right),
\end{equation}
where $P^S_{su,n}$ is the BS transmitted power allocated to RB $n$, $\mathcal{N}_0$ is the noise power, and $\mathcal{I}^S_{u}$ is the inter-cell interference at the user $u$ caused by closest BS (no intra-cell interference on the downlink direction between different tiers is assumed) and expressed as follows:
\begin{equation}\label{interference}
  \mathcal{I}^S_{su}=\hspace{-.1cm}\left\{
   \begin{array}{ll}
   \hspace{-.2cm}\sum\limits_{\substack{\tilde{m}=1 \\ \tilde{m}\neq m}}\limits^{M}\left(\sum\limits_{\substack{\tilde{u}=1 \\ \tilde{u}\neq u}}^{U}\epsilon^M_{\tilde{m}\tilde{u},n}  P^M_{\tilde{m}\tilde{u},n} \right) h^M_{\tilde{m}u,n}, \hspace{-.3cm}& \hbox{for ground tier,} \\
   \hspace{-.2cm}0, \hspace{-.3cm}& \hbox{for air tier,} \\
   \hspace{-.2cm}0, \hspace{-.3cm}& \hbox{for space tier,}
                   \end{array}
                 \right.
\end{equation}
where $\epsilon^M_{\tilde{m}\tilde{u},n}$ is representing the exclusivity of the TBS $m$ and RB allocation: $\epsilon^M_{\tilde{m}\tilde{u},n}=1$, if RB $n$ of nearby station is allocated to another user $\tilde{u}$ from TBS $m$, and $\epsilon^S_{\tilde{s}\tilde{u},n}=0$, otherwise. In fact, since the same RB might be allocated to the more than one user in different TBSs simultaneously in the ground tier, an inter-cell interference might be caused to some users.
In~\eqref{interference}, it can be noticed that the inter-cell interference is 0 for both air and space tier since we ignored this interference because the HAPs and satellite station are assumed to be equipped with different antennas and different beams that allow managing the resources in an efficient way by using different FSs for different ground cells~\cite{HAP5}.
Therefore, the BH data rate from station $w$ to HAP $l$ can be expressed as:
\begin{equation}
\bar{R}_{wl}=\delta_{wl} B_0 \log_2\left(1+\frac{P_0 g_{wl}}{\mathcal{N}_0 B_0} \right),
\end{equation}
where $B_0$ and $P_0$ are the BH bandwidth and transmit power at station $w$, respectively.

\section{Problem Formulation}\label{ProbelmFormulation}
Our goal is to maximize the downlink data rate utility of ground users by optimizing the FH and BH associations, BSs' transmit power, and HAPs' locations.
Based on the system parameters, the optimizer will determine whether using satellite station or HAPs would be more efficient than using TBSs.
For remote areas or heavy loaded areas where the TBSs' would not able to support all ground users, the optimizer may decide to use the HAPs and satellite station when needed to support the ground users. This depends on many factors such as the users' location, number of the users to be served, and existence of the TBSs. For instance, if the users in remote areas outside the coverage of TBSs, then the satellite station and HAPs will try to accommodate these users. In the case of highly loaded networks, the optimizer is forced to use the network full capacity including the HAPs and satellite station in order to meet the users' demand. The optimization problem that we formulate will also optimize the locationsof the HAPs and the BH association such that the throughput is maximized.
Therefore, our optimization problem can be formulated as follows:

\begin{align}
&\hspace{-0.5cm}\underset{\substack{\epsilon^S_{su,n},\delta_{wl}\\ P^S_{su,n}, \bold{J}^L_l}}{\text{maximize}} \quad \mathcal{U}(R_u)\label{of}\\
&\hspace{-0.5cm}\text{subject to:}\nonumber\\
%&\hspace{-0.5cm}  0 \leq \sum\limits_{n=1}^{N^M}\sum\limits^U_{u=1}  P^M_{mu,n} \leq \bar{P}_M, \quad \forall m, \label{powerm}\\
%&\hspace{-0.5cm}  0 \leq \sum\limits_{n=1}^{N^L}\sum\limits^U_{u=1}  P^L_{lu,n} \leq \bar{P}_L, \quad \forall l, \label{powerl}\\
&\hspace{-0.5cm}  0 \leq \sum\limits_{n=1}^{N^S}\sum\limits^U_{u=1} \epsilon^S_{su,n} P^S_{su,n} \leq \bar{P}_S, \quad \forall s, S \in\{M,L,0\} \label{powerm}\\
%&\hspace{-0.5cm}  0 \leq \sum\limits_{n=1}^{N}\sum\limits^U_{u=1} \epsilon_{0u,n} P_{0u,n} \leq \bar{P}_0, \label{power0}\\
%&\hspace{-0.5cm}   \sum\limits_{l=1}^{L} \delta_{gl} \sum\limits_{n=1}^{N^L} \sum\limits^U_{u=1} R^L_{lu,n}  \leq \bar{R}_g(\delta_{gl}), \quad \forall g, \label{rate_backhauling}\\
&\hspace{-0.5cm}   \sum\limits_{n=1}^{N^L} \sum\limits^U_{u=1} R^L_{lu,n}  \leq \sum\limits_{w=0}^{W} \bar{R}_{wl}, \quad \forall l, \label{rate_backhauling}\\
%&\hspace{-0.5cm}  \sum_{n=1}^{N^M}\epsilon^M_{mu,n} + \sum_{n=1}^{N^L}\epsilon^L_{lu,n} + \sum_{n=1}^{N}\epsilon_{0u,n} \leq 1, \quad \forall m, \forall l, \forall u,\label{asso}\\
&\hspace{-0.5cm} \sum_{S \in\{M,L,0\}} \sum_{s=1}^{S} \sum_{n=1}^{N^S}\epsilon^S_{su,n} \leq 1, \quad \forall u, \label{asso}\\
&\hspace{-0.5cm} \sum_{u=1}^{U}\epsilon^S_{su,n} \leq 1, \quad \forall s, \forall S, \forall n=1,..,N^S, \label{RB_BS}\\
%&\hspace{-0.5cm} \sum_{u=1}^{U}\epsilon_{0u,n} \leq 1, \quad \forall n=1,..,N, \label{RB_satellite}\\
&\hspace{-0.5cm}  \sum\limits_{w=0}^W \delta_{wl} = 1, \quad \forall l,\label{asso_backhauling1}\\
&\hspace{-0.5cm}   \sum\limits_{l=0}^L \delta_{wl} \leq  \bar{L}_w \quad \forall w,\label{asso_backhauling2}
\end{align}
where $\mathcal{U}(R_u)$ denotes the data rate utility of all users. Constraint~\eqref{powerm} represents the peak power constraints at BS $s$. Constraint~\eqref{rate_backhauling} represents the BH constraint of HAPs. Note that, we assume that the BH links of TBSs using fiber links and satellite station using feeder link are perfect, and hence not considering them in the BH constraint.
Constraint~\eqref{asso} is to ensure that each ground user can be associated with one BS over one RB.
Constraint~\eqref{RB_BS} is to ensure there are no intra-cell interference between users associated to the same BS.
Finally, constraints~\eqref{asso_backhauling1} and~\eqref{asso_backhauling2} represent the BH association conditions of HAPs.

\subsection{Utility Selection}
\label{utilities}
In this section, we characterize two different utility metrics that will be employed in the optimization problem given in~\eqref{of}-\eqref{asso_backhauling2}.

\subsubsection{\textbf{Max-Sum Utility (MSU)}}
The utility of this metric is equivalent to the sum FH data rate of the ground users $\mathcal{U}(R_u)=\sum^{U}_{u=1} R_u$ as it promotes users with favorable channel gains and interference by allocating to them the best resources. On the other hand, using this utility will deprive users with bad channel gains and interference from the resources and thus they will have very low data rates~\cite{max_rate}.
\\
\subsubsection{\textbf{Max-Min Utility (MMU)}}
Due to the unfairness of MSU resource allocation, the need for more fair utility metrics arises. Max-min utility (MMU) is attempting to maximize the minimum data rate in the total network such as $\mathcal{U}(R_u)=\underset{u}{\min}(R_u)$~\cite{Min-Max}
Note that, by increasing the priority of users having lower rates, MMU leads to more fairness in the network. In order to simplify the problem for this approach, we introduce a new decision variable $R_{\min}=\underset{u}{\min}(R_u)$. Therefore, our optimization problem can be re-formulated as:
\begin{align}
&\hspace{-0.5cm}\underset{\substack{\epsilon^S_{su,n},\delta_{wl}\\ P^S_{su,n}, \bold{J}^L_l, R_{\min}}}{\text{maximize}} \quad R_{\min} \label{ofmmu}\\
&\hspace{-0.5cm}\text{subject to:}\nonumber\\
  &\hspace{-0.5cm} R_u  \geq R_{\min}, \quad \forall u, \label{minConst}\\
 &\hspace{-0.5cm} \eqref{powerm},\eqref{rate_backhauling},\eqref{asso},\eqref{RB_BS},\eqref{asso_backhauling1}, \eqref{asso_backhauling2}.\nonumber
 \end{align}
%\subsubsection{\textbf{Proportional Fair Utility (PFU)}}
%A tradeoff between the maximization of the sum rate and the maximization of the minimum rate could be the maximization of the geometric mean data rate $U(R_u)=(\prod^{U}_{u=1} R_u)$, which is equivalent to $U(R_u)=\sum^{U}_{u=1} \ln(R_u)$~\cite{PF}. The proportional fair utility (PFU) metric is fair, since a user with a data rate close to zero will make the whole product go to zero. Hence, any algorithm maximizing the geometric means would avoid having any user with very low data rate. In addition to this, the metric will reasonably promote users with good wireless channels (capable of achieving high data rates), since a high data rate will contribute in increasing the product.

The formulated optimization problem in~\eqref{of}-\eqref{asso_backhauling2} is a mixed integer non-linear programming (MINLP), and solving it is a challenging task. In order to simplify the problem, we propose to solve it in two stages: short-term stage and long-term stage. In the former stage, we propose two solutions (near optimal and low complexity solutions) to find the best associations and transmit power allocations. While in the latter stage, the problem of optimizing the locations of HAPs is investigated.

\section{Short-term Stage}\label{solutions}

In the short-term stage and considering fix HAPs' locations, we firstly solve for the FH associations (i.e, $\epsilon^S_{su,n}$) and BH associations (i.e., $\delta_{wl}$) using uninform power distributions, i.e., $P^S_{su,n}=\bar{P}_S/N^S$. We then optimize the transmit powers of the BSs for these associations by approximating the optimization problem using Tyler expansion. %Finally, given these associations and powers, we propose an efficient algorithm to optimize the locations of the HAPs' in order to improve the system throughput.
%Note that, even fox fix associations and HAPs' location, the proble is still non convex in terms of $P^S_{su,n}$. Therefore, we propose two solutions to solve our optimization problem. The first solution achieves near optimal solution based on dual decomposition method. In this solution, we are solving the problem in iterative way for the transmit power of the  subgradient method\cite{subgradient}.

In the sequel, we propose two solutions in this stage. In the first solution, we propose a near optimal solution based on Taylor expansion approximation. While in the second solution, an low complexity heuristic approach using frequency partitioning (FP) technique is proposed, where we omit the inter-cell interference among TBSs and, thus, solve the problem quickly.

\subsection{Near Optimal Solution}
%Because of the large value of the PL between the satellite station and ground users, we assume that the downlink communications from satellite station to ground users will be provided when needed (i.e., when the network that includes TBSs and HAPs fails to sever the ground users). This failing can be either because the users are out of the coverage areas of TBSs and HAPs or because the congestion of the network (e.g., large number of users). Therefore, $\epsilon_{0u,n}$ and $P_{0u,n}$ can be optimized after solving the main fronthauling problem that optimizes the following variables: $\epsilon^M_{mu,n}, \epsilon^L_{lu,n}, P^M_{mu,n}, P^L_{lu,n}$ and $R_{\min}$.
%The optimization problem of the main fronthauling problem can be given as:
%\begin{align}
%&\hspace{-0.7cm}\textbf{(\text{P2}):} \underset{\substack{\epsilon^M_{mu,n}, \epsilon^L_{lu,n}, \\ P^M_{mu,n}, P^L_{lu,n}, R_{\min}}}{\text{maximize}} \quad R_{\min}\label{ofm}\\
%&\hspace{-0.5cm}\text{subject to:}\nonumber\\
% &\hspace{-0.5cm} \eqref{powerm}, \eqref{rate_backhauling}, \eqref{asso},\eqref{RB_BS}, \eqref{min2}, \eqref{min3}. \label{other_constraints2}
% \end{align}

In fact, by fixing the transmit powers, it can be seen that the optimization problem becomes linear binary optimization problem in terms of $\epsilon^S_{su,n}$ and $\delta_{wl}$, except~\eqref{of} which is non linear with respect to $\epsilon^S_{su,n}$ due to the existence of the $\mathcal{I}^M_{mu}$ term in~\eqref{interference}. On the other hand, by fixing $\epsilon^S_{su,n}$ and $\delta_{wl}$, the optimization problem is convex problem in $P^S_{su,n}$ except constraints~\eqref{of} and~\eqref{rate_backhauling}.
Therefore, in order to solve the problem iteratively (solving the associations and them transmit powers), the goal is to linearize~\eqref{of} respects to $\epsilon^S_{su,n}$ with fixed $P^S_{su,n}$. Then, approximate~\eqref{of} and~\eqref{rate_backhauling} to convex ones respects to  $P^S_{su,n}$ with fixed $\epsilon^S_{su,n}$ and $\delta_{wl}$.

It can be notice that $\mathcal{I}^M_{mu}$ in~\eqref{interference} can be re-write as $\sum\limits_{\substack{\tilde{m}=1 \\ \tilde{m}\neq m}}\limits^{M}\left(\sum\limits_{\substack{\tilde{u}=1 \\ \tilde{u}\neq u}}^{U} P^M_{\tilde{m}\tilde{u},n} \right) h^M_{\tilde{m}u,n}$ by adding the following constraint:
\begin{equation}\label{add}
 0 \leq P^M_{\tilde{m}\tilde{u},n} \leq \epsilon^M_{\tilde{m}\tilde{u},n}  \bar{P}_M, \quad \forall \tilde{m}, \forall \tilde{u}.
\end{equation}
Now, the optimization problem becomes a linear one with respect to $\epsilon^S_{su,n}$ and $\delta_{wl}$ and thus, it can be solved using on-the-shelf softwares such as Gurobi/CVX interface~\cite{Gurobi}.

On the other hand, by fixing the association variables (i.e., $\epsilon^S_{su,n}$ and $\delta_{wl}$), the optimization problem in~\eqref{of}-\eqref{asso_backhauling2} can be re-write as:
\begin{align}
&\hspace{-0.5cm}\underset{P^S_{su,n}}{\text{maximize}} \quad \mathcal{U}(R_u)\label{of1}\\
&\hspace{-0.5cm}\text{subject to:}\nonumber\\
&\hspace{-0.5cm}  0 \leq \sum\limits_{n=1}^{N^S}\sum\limits^U_{u=1} \epsilon^S_{su,n} P^S_{su,n} \leq \bar{P}_S, \quad \forall s, S \in\{M,L,0\} \label{powerm1}\\
&\hspace{-0.5cm}   \sum\limits_{n=1}^{N^L} \sum\limits^U_{u=1} R^L_{lu,n}  \leq \sum\limits_{w=0}^{W} \bar{R}_{wl}, \quad \forall l. \label{rate_backhauling1}
\end{align}

It can be noticed that the optimization problem~\eqref{of1}-~\eqref{rate_backhauling1} is convex problem in $P^S_{su,n}$ except constraints~\eqref{of1} and~\eqref{rate_backhauling1}.
Therefore, the goal is to convert these constraints into convex ones in terms of $P^S_{su,n}$ in order to solve the power optimization problem efficiently.
%
%
%Therefore, we firstly start by introducing a new variable $\rho^S_{su,n}$ for each link to linearize the product of the transmit powers and users' association such as $\rho^S_{su,n}=\epsilon^S_{su,n} P^S_{su,n}$ where the following inequalities have to be respected:
%\begin{align}
%&P^S_{su,n} \geq \rho^S_{su,n} \geq 0, \quad \forall s, u, n \label{linear1}\\
%&\rho^S_{su,n} \geq \bar{P}_S \epsilon^S_{su,n} - \bar{P}_S + P^S_{su,n}, \quad \forall s, u, n \label{linear2}\\
%&\rho^S_{su,n} \leq \bar{P}_S \epsilon^S_{su,n}, \quad \forall s, u, n.\label{linear3}
%\end{align}
%The first inequality ensures that $\rho^S_{su,n}$ is between 0 and $P^S_{su,n}$. The second and third inequalities guarantee
%that $\rho^S_{su,n}=0$ if $\epsilon^S_{su,n}=0$, and $\rho^S_{su,n}=P^S_{su,n}$ if $\epsilon^S_{su,n}=1$.
%The third inequality also guarantees that $\rho^S_{su,n}$ cannot exceed $\bar{P}_S$.

\subsubsection{Approximation of The Objective Function~\eqref{of1}}: Let us start with objective function~\eqref{of1} by expanding it as follows:
\begin{align}\label{power_non0}
R_u=&B^M\sum_{m=1}^{M} \sum_{n=1}^{N^M} \epsilon^M_{mu,n} \log_2\left(1+\frac{P^M_{mu,n}\,h^M_{mu,n}}{\mathcal{I}^M_{u}+\mathcal{N}_0 B^M}\right) + \nonumber \\&
B^L \sum_{l=1}^{L} \sum_{n=1}^{N^L} \epsilon^L_{lu,n} \log_2\left(1+\frac{P^L_{lu,n}\,h^L_{lu,n}}{\mathcal{N}_0 B^L}\right)  + \nonumber \\&
B^0 \sum_{n=1}^{N^0} \epsilon^0_{0u,n} \log_2\left(1+\frac{P^0_{0u,n}\,h^0_{0u,n}}{\mathcal{N}_0 B^0}\right)  .
\end{align}
In~\eqref{power_non0}, since we are maximizing the utility, then we need the function to be concave. Note that the second and third terms are concave functions in terms of $P^L_{lu,n}$ and $P^0_{0u,n}$. In order to convert the objective function to a concave function, the first term in~\eqref{power_non0} needs to be concave it terms of $P^M_{mu,n}$.
Let us start by expanding the first term in~\eqref{power_non0} as follows:
\small
\begin{align}\label{power_non1}
&R^M_{mu,n}
 =\underbrace{\epsilon^M_{mu,n} B^S \log_2\left(\mathcal{N}_0 B^S  + \sum\limits_{m=1}^{M}\sum\limits_{u=1}^{U} P^M_{mu,n} h^M_{mu,n} \right)}_{\hat{R}^M_{mu,n}}
\nonumber \\&
\underbrace{- \epsilon^M_{mu,n} B^S \log_2\left(\mathcal{N}_0 B^S + \sum_{\substack{\tilde{m}=1 \\ \tilde{m}\neq l}}\limits^{M}\sum_{\substack{\tilde{u}=1 \\ \tilde{u}\neq u}}^{U}  \epsilon^M_{\tilde{m}\tilde{u},n} P^M_{\tilde{m}\tilde{u},n} h^M_{\tilde{m}u,n} \right).}_{\check{R}^M_{mu,n}}
\end{align}
\normalsize
Note that $\hat{R}^M_{mu,n}$ is concave, because the $\log$ of an affine function is concave~\cite{Boyd}. Also, $\check{R}^M_{mu,n}$ is a convex function, and thus, it needs to be converted to a concave function. To tackle the non-concavity of $\check{R}^M_{mu,n}$, the successive convex approximation (SCA) technique can be applied where in each iteration, the original function is approximated by a more tractable function at a given local point as given in algorithm~\ref{SCA Algorithm}. Recall that $\check{R}^M_{mu,n}$ is convex in $P^M_{\tilde{m}\tilde{u},n}$, and any convex function can be globally lower-bounded by its first order Taylor expansion at any point. Therefore, given $P^M_{\tilde{m}\tilde{u},n}(r)$ in iteration $r$, we obtain the following lower bound for $\check{R}^M_{mu,n}(r)$:
\begin{align}\label{power_con}
\check{R}^M_{mu,n}(r) \geq & - \epsilon^M_{mu,n} B^S \log_2\left(\psi(r) \right) \nonumber \\& -\frac{\epsilon^M_{\tilde{m}\tilde{u},n} h^M_{\tilde{m}u,n}}{\ln(2)\psi(r)}(P^M_{\tilde{m}\tilde{u},n} -P^M_{\tilde{l}\tilde{u},n}(r)),
\end{align}
where $\psi(r)=\mathcal{N}_0 B^S + \sum_{\substack{\tilde{m}=1 \\ \tilde{m}\neq l}}^{M}\sum_{\substack{\tilde{u}=1 \\ \tilde{u}\neq u}}^{U}  \epsilon^M_{\tilde{m}\tilde{u},n} P^M_{\tilde{m}\tilde{u},n} h^M_{\tilde{m}u,n} $.
		\begin{algorithm}[t]
			\caption{SCA Algorithm}
			\label{SCA Algorithm}
			\begin{algorithmic}[l]
				\STATE Select feasible initial values $P^M_{\tilde{m}\tilde{u},n}$.
                \STATE r=1.				
                \REPEAT
				\STATE Solve the optimization problem~\eqref{of1}-\eqref{rate_backhauling1} using the interior-point method to determine the new approximated solution $P^M_{\tilde{m}\tilde{u},n}(r)$.
				\UNTIL Convergence.
			\end{algorithmic}
		\end{algorithm}
Now it can be seen that the objective function in~\eqref{power_non0} is a concave function in terms of $P^S_{su,n}$.
%Note that, the same technique can be used for constraint.
%Therefore, in the sequel, we will explain the details of how to convert constraint~\eqref{rate_backhauling} to a convex one as example.

\subsubsection{Approximation of Constraint~\eqref{rate_backhauling1}}:
We can approximate this constraint by ensuring that the data rate from HAP $l$ to any associated users $u$ is smaller that the average BH constraint. In other words, constraint~\eqref{rate_backhauling} can be approximated as:
\begin{equation}\label{back_appro}
 \sum\limits_{n=1}^{N^L} \epsilon^L_{lu,n} \left(1+\frac{P^L_{lu,n}\,h^L_{lu,n}}{\mathcal{N}_0 B^S} \right)  \leq \frac{\omega}{\sum\limits_{n=1}^{N^L} \epsilon^L_{lu,n} }, \quad \forall l, \forall u,
\end{equation}
where $\omega=2^{(\sum_{w=0}^{W} \bar{R}_{wl})/B^S}$.

Now, our power allocation optimization problem given in~\eqref{of1}-\eqref{rate_backhauling1} becomes convex problem in respect to $P^S_{su,n}$ for fix associations and HAPs' locations. Thus, the duality gap of our convex power allocation problem in OFDMA system is zero. Hence, We can solve our convex transmit power allocation optimization problem by exploiting its strong duality.

\subsection{Low Complexity Solution}
\begin{figure}[h!]
  \centerline{\includegraphics[width=2.5in]{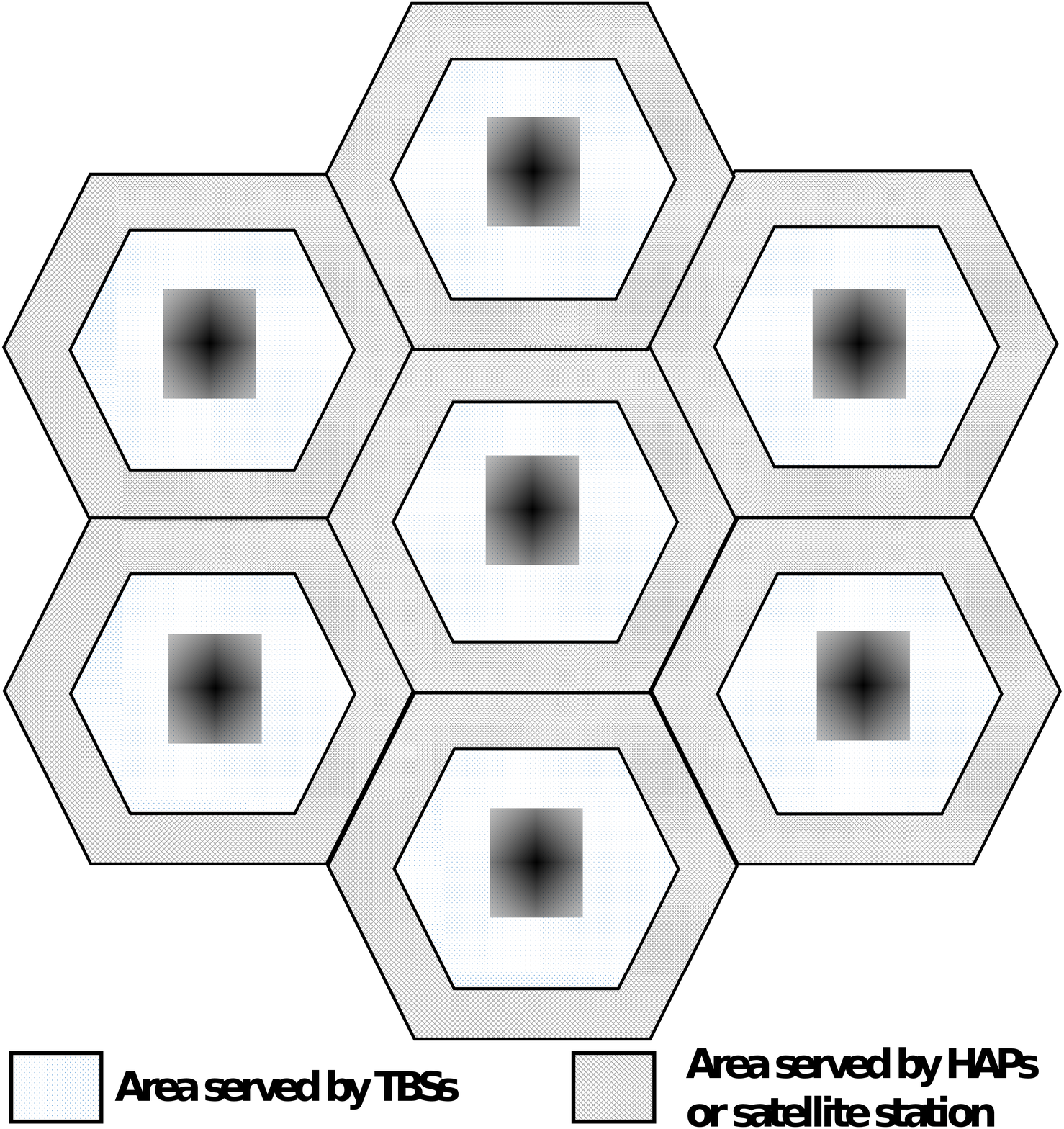}}\vspace{-0.2cm}
   \caption{Frequency partitioning technique.}\label{FP}
\end{figure}

In subsection, we propose a low complexity and efficient technique to mitigate the inter-cell interference between users associate
to different TBSs in the ground tier. We divide the users into two groups, 1) center users: users located in areas close to the center of the TBSs' cells and 2) edge users: users located in the edges of the TBSs' cells.
We assume that TBSs can be associated with the center users only. In this case, inter-cell interference is assumed to be $\mathcal{I}^M_{mu}=0$ due to large distances between users that using the same frequency bands in the ground tier. Hence, the association problem is linear and we can omit constraint~\eqref{add}. On the other hand, the edge users can be served by the HAPs and satellite station as shown in Fig~\ref{FP}. Hence,~\eqref{of1} becomes a convex function with respect to $P^S_{su,n}$.

Therefore, the formulated optimization problem in~\eqref{of}-~\eqref{rate_backhauling} is a linear assignment problem with respect to $\epsilon^S_{su,n}$ and $\delta_{wl}$ and can be solved efficiently by using the Hungarian algorithm with complexity of $(M N^M+L N^L + N^0)^3$~\cite{hungarian_method}.

On the other hand, by only approximating~\eqref{rate_backhauling1}, the formulated optimization problem in~\eqref{of1}-~\eqref{rate_backhauling1} becomes a convex optimization problem in terms of $P^S_{su,n}$. Therefore, it can be solved efferently by solving the its dual problem.

\section{Long-term Stage}\label{solutionsHAP}
%We consider optimizing the placement of the UAVs for given associations and UAVs' transmit power levels.

Due to the non-convexity of the formulated optimization problem~\eqref{of}-\eqref{asso_backhauling2} even with fixed associations and BSs transmit powers, we introduce a low complexity and efficient algorithm based on the SR process. The proposed algorithm has many advantages over other heuristic algorithms proposed in the literature such as simple implementation, low complexity, and quick convergence to a near optimal solution.
Furthermore, since we solve this problem for long-term stage, where the time slots are relatively long compared to the channel coherence time and hence, we focus on the average statistics (such as average channel gains) of the network. This implies that average users' throughput is assumed instead of instantaneous throughput.

\begin{figure}[h!]
  \centerline{\includegraphics[width=2.5in]{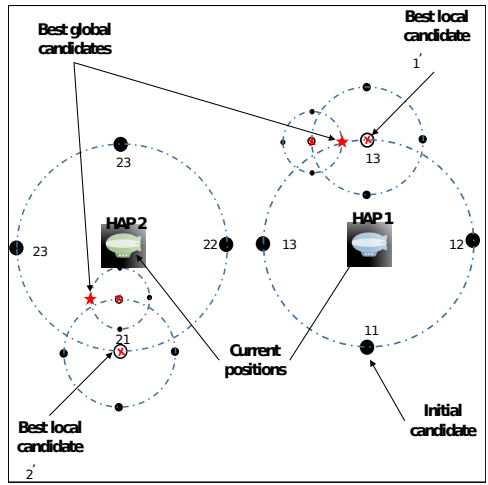}}
   \caption{Proposed heuristic approach.}\label{SRfig}
\end{figure}

We start our algorithm by generating initial next position candidates $T_l, \forall l$ as a circle with radius $r(i)$ around all current locations of the HAPs to form the inial population.
Next, we solve the average associations and transmit powers of the BSs to determine maximum number of average users inside the HAPs coverage areas for each candidates combination.
In other words, determine the value $U^L_{\max}=\sum^L_{l=1} \sum^{N^S}_{n=1} \sum^U_{u=1} \epsilon^L_{lu,n}$, after excluding the users associated to TBSs.
We then find the initial best local candidate combinations $T^{i,\text{local}}=t_l^{i,\text{local}}, \forall l$ that gives the maximum number of users associated with HAPs for iteration $i$.
After that, we apply the SR process recursively to find the best global solution $T^*=t_l^*, \forall l$ by generating a new candidates on a circle of radius ($\textit{r(i+1)=r(i)/2}$) around each local solution. We repeat this process until the size of the sample space decreases
below a certain threshold $I_{\max}$ or no improvement can be made. Fig.~\ref{SRfig} shows an example of the proposed algorithm using two HAPs ($L=2$), $T=4$, and three maximum iteration $I_{\max}=3$.
The details of the HAPs' locations algorithm that optimize the locations of the HAPs based on average users' locations are given in Algorithm~\ref{joint}.

\begin{algorithm}[h!]
\caption{HAPs' Locations Algorithm}\label{joint}
\small
\begin{algorithmic}[1]
\STATE i=1.
\STATE Generate initial candidates $T$ in a circle of radius $r=r(i)$ around each HAP
$J_l^{\mathcal{L}}(t_l),\;l=1 \cdots L$.
\WHILE {{Not} converged or reaching $I_{\max}$}
\FOR {$l=1 \cdots L$}
\FOR {$t_l=1 \cdots T$}
\STATE Find the optimized value of $\epsilon^S_{su,n}$, $\delta_{wl}$, and $P^S_{su,n}$ as explained in Section. IV for average statistics.
\STATE Compute $U^L_{\max}=\sum^L_{l=1} \sum^{N^S}_{n=1} \sum^U_{u=1} \epsilon^L_{lu,n}$.
\ENDFOR
\ENDFOR
\STATE Find $t_l^{i,\text{local}}=\underset{t_t}{\arg\mathrm{max}}\, U^L_{\max}$, (i.e., $t_l^{i,\text{local}}$ indicates the index of the best local candidate that results in the highest $U^L_{\max}$ for iteration $i$).
\STATE $r=r(i)/2$
\STATE Start applying SR process for the local solution.
\STATE i=i+1.
\ENDWHILE
\end{algorithmic}
\normalsize
\end{algorithm}

\section{Simulation Results}\label{Simulation}
In this section, selected simulation results are provided to demonstrate the advantages of our proposed solution. The simulation results are set within area of  180 Km $\times$ 180 Km. Within this area, $U$ users are distributed in three different subareas (i.e., subarea 1: 30 Km$^2$, subarea 2: 30 Km$^2$, and subarea 3: the remaining) with different dense distributions. Subarea 1 contains $M=9$ TBSs with coordinates are: x: (75 Km to 105 Km), and y: (0 Km to 30 Km) and contains 40\% of the total number of users. Subarea 2, has no TBSs with coordinates are:  x: (75 Km to 105 Km), and y: (150 Km to 180 Km) and contains 30\% of the total number of users. In other words, subarea 2 has dense users but with no infrastructure (e.g., it ca flying at a fixed altitude be an area with a damaged infrastructure or temporarily high traffic area). While subarea 3 is the remaining area and contains 30\% of the total number of users. For example, subarea 1 and subarea 2 can be considered as urban areas with and without TBSs, respectively, while subarea 3 can be considered as rural are.
The satellite station is assumed to be in a fixed location with the coordinate [90,90,2000] Km. Also, we consider $L=5$ HAPs flying at a fixed altitude $z^L_l=18$ Km, $\forall l=1,..,L$. We assume that the HAPs initially start at location $\bold{J}^L_1=[90,90,18]$ Km, $\bold{J}^L_1=[30,30,18]$ Km, $\bold{J}^L_1=[150,30,18]$ Km, $\bold{J}^L_1=[30,150,18]$ Km, $\bold{J}^L_1=[150,150,18]$ Km. We assume that $W=4$ and located in the four corners of the desired area.
We assume $N^M=50, N^L=100$, and $N^0=200$ available RBs in the FH in our simulations. The maximum transmit power of TBS, HAP, and satellite station are, respectively, given as \{40,100,250\} W. The noise power $N_0$ is assumed to be $-174$ dBm.
In Table~\ref{tab2}, we present the values of the remaining environmental parameters used in the simulations unless otherwise stated~\cite{HAP3,HAP_channel2,Sat_channel}.

{\small
\begin{table}[b]
\centering
\caption{\label{tab2} Simulation parameters}
\addtolength{\tabcolsep}{-2pt}\begin{tabular}{|l|c||l|c||l|c|}
\hline
\textbf{Constant} & \textbf{Value} & \textbf{Constant} & \textbf{Value}& \textbf{Constant} & \textbf{Value}\\ \hline \hline
$f_c^M$ (GHz) & 1.8 & $f_c^L$ (GHz) & $3.0$ & $f_c^0$ (GHz) & $5.0$   \\ \hline
$f_c$ (GHz) & $3.4$ & $\chi$ & $2$ & $\kappa_L$,$\kappa_0$ & $10$ \\ \hline
$B^M$ (kHz) & 1.8 & $B^L$ (MHz) & $1.0$ & $B^L$ (MHz) & $2.0$   \\ \hline
$\Omega_0$ & $0.372$ & $\Omega_1$ & $0.0129$ & $\Omega_2$ & $7.64$ \\ \hline
\end{tabular}
\end{table}
}

\begin{figure}[h!]
  \centerline{\includegraphics[width=3.4in]{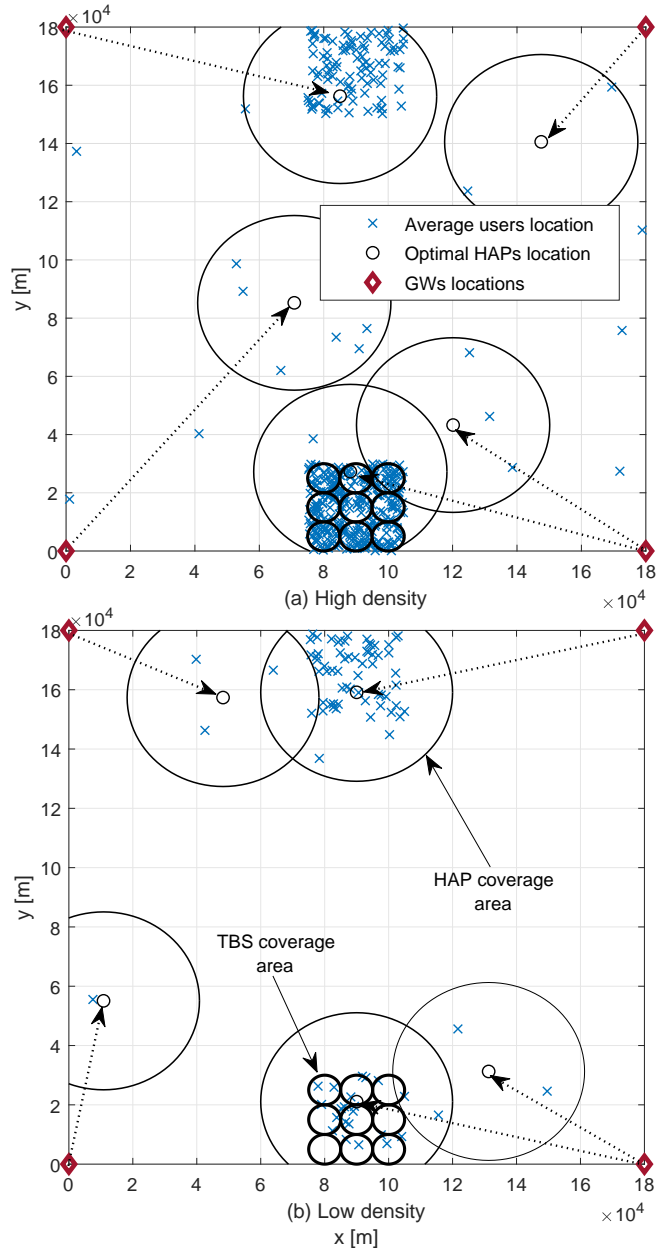}}
   \caption{HAP placement optimization with BH associations.}\label{HAP_location}
\end{figure}

Fig.~\ref{HAP_location} plots the optimization of HAPs' locations and BH associations using average users location for two high and low users densities with $P_0=40$ W, $B_0=4$ MHz, and $\bar{P}_l=100$ W. The figure also shows the BH association of HAPs with the GWs. For instance, Fig.~\ref{HAP_location}-a shows that the HAPs try to accommodate maximum number of users by finding the best HAPs' locations that improved the FH and BH links at the same time. Note that the users located outside the coverage areas of HAPs and TBSs are associated with satellite station. While, in the case of low dense users, HAPs try to associated with all users as shown in Fig.~\ref{HAP_location}-b. This is can explained by the fact that associating with HAPs, in general, provide a better data rate instead of associating to satellite station due to the high path-loss attenuation between satellite station and ground users. In this case, the HAPs can provide services to ground users with much lower delay and higher throughput.

\begin{figure}[h!]
  \centerline{\includegraphics[width=3.4in]{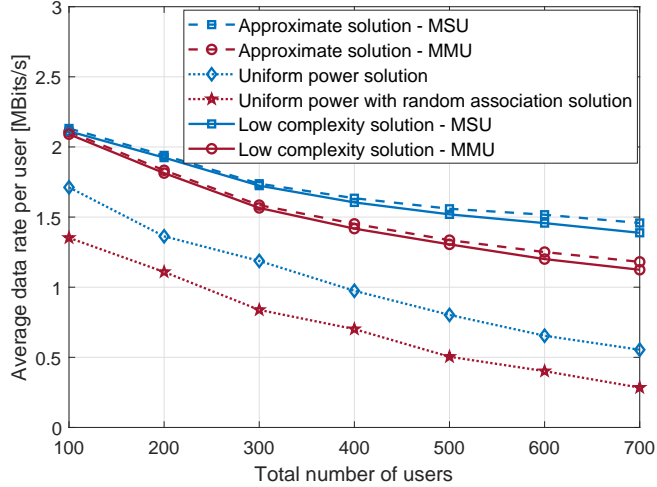}}
   \caption{Average data rate per user as a function of number of users.}\label{fig1}
\end{figure}

Fig.~\ref{fig1} plots the average data rate per user (i.e., $\sum^U_{u=1}$/$U$) versus total number of users (i.e., subarea 1: 40\%, subarea 2: 30\%, subarea 3: 30\%) with $P_0=40$ W, $B_0=4$ MHz, and $\bar{P}_l=100$ W. Our proposed solutions (i.e., approximated and low complexity solutions) are compared with two benchmark solutions:
1- optimizing only the FH and BH associations and the HAPs' locations with uniform power distribution (i.e., $P^S_{su,n}=\bar{P}_S/N^S$),
and 2- optimizing only the placement of the HAPs using random FH and BH associations with uniform power distribution.
The figure shows that for fixed resources, as $U$ increases, the average data rate decreases due to the limitation of the available resources.
Furthermore, the figure shows that the approximate solution and low complexity solution achieve almost the same performance for low values of $U$. However, there will be a gap between the two proposed solutions when $U$ is relatively large. This is can be explained by the fact that the low complexity solution forces some HAPs to cover the TBSs coverage areas in order to implement the frequency partitioning technique as shown in Fig.~\ref{FP}. This will not effect the performance when $U$ is relatively low, but when $U$ is large, and because of the limitation number of HAPs, then the HAPs located in the TBSs coverage areas maybe be better to move to different locations. Thanks to the HAPs, because HAPs have large coverage areas of HAP, this gap for large $U$ is still acceptable.
This figure also shows that our proposed solutions outperform the other two benchmarks solutions. For instance, using $U=400$, our proposed solution can enhance the average rate throughput by at least 39\% and 88\% compared to optimize the associations with uniform power and to random associations with uniform power, respectively.
Furthermore, it can be noticed that the gap between the proposed solutions and benchmarks solutions is increased as $U$ increases. This is due to the fact, as $U$ increases, the need of managing and optimizing the power is needed more.

All simulations show that MSU leads to the highest average data rate in the system. However, this comes at the expense of fairness as it is shown in Table~\ref{Tabbig}. Indeed, the table compares between the two different utilities for the same channel realization with fixed $P_0=40$ W, $B_0=4$ MHz, $\bar{P}_l=100$ W, and $U=400$ (for this realization, 160 user associated to TBSs, 172 user associated to HAPs 172, and 68 user associated to satellite station). In Table~\ref{Tabbig}, we denote $\bar{R}^S$, $R^S_{\max}$, and  $R^S_{\min}$ as average rate, maximum rate, and minimum rate in tier $S$, respectively. Also, $P^S_{\max}$ and $P^S_{\min}$ denoted as maximum and minimum transmit powers in tier $S$.
By using one realization, it can be shown that MSU enhances the average data rate, by allocating most of the resources to users having the best channel conditions. On the other hand, MMU approach maximizes the minimum data rate and provides almost the same rate for all users. Hence, MMU leads to more fairness performance. The choice of the utility is related to the service given to the users. For instance, if the application requires same downlink rates betweenisers, then MMU can be used. on the other hand, if it consists in a pure transmission without priorities, then MSU could be employed.

\begin{table}[t!]
\begin{center}
\caption{Comparison between MSU and MMU for fixed $P_0=40$ W, $B_0=4$ MHz, $\bar{P}_l=100$ W, and $U=400$ }
\label{Tabbig}
\begin{tabular}{|c|c|c|c|c|}%
\hline
   \textbf{} & Users associated  & $\bar{R}^S$  & ($R^S_{\max},P^S_{\max}$) & ($R^S_{\min},P^S_{\min}$) \\
  %\cline{3-5}
  \textbf{}  & with   & Mbits/s & (Mbits/s,W) &(Mbits/s,W) \\
  \cline{1-5}
  \multirow{3}{*}{\begin{turn}{90}\textbf{MSU}\end{turn}}  & TBSs   & $2.14$ & $(4.13,15.25)$ & $(\sim 0,\sim 0)$\\
  \cline{2-5}
  \textbf{}  & HAPs   & $1.92$ & $(3.19,39.70)$ & $(.0415,1.2)$ \\
  \cline{2-5}
  \textbf{}  & Satellite   &  $(0.016)$ & $(0.085,97.53)$ & $(\sim 0,\sim 0)$ \\
\hline\hline
    \multirow{3}{*}{\begin{turn}{90}\textbf{MMU}\end{turn}}  & TBSs   & $(1.91)$ & $(1.96,5.49)$ & $(1.88,8.20)$\\
  \cline{2-5}
  \textbf{}  & HAPs   & $(1.50)$ & $(1.81,3.91)$ & $(1.39,4.5)$ \\
  \cline{2-5}
  \textbf{}  & Satellite   & $(0.01)$ & $(0.01,4.81)$ & $(0.01,4.65)$ \\
  \hline
  \end{tabular}
\end{center} \vspace{-0.5cm}
\end{table}

\begin{figure}[h!]
  \centerline{\includegraphics[width=3.4in]{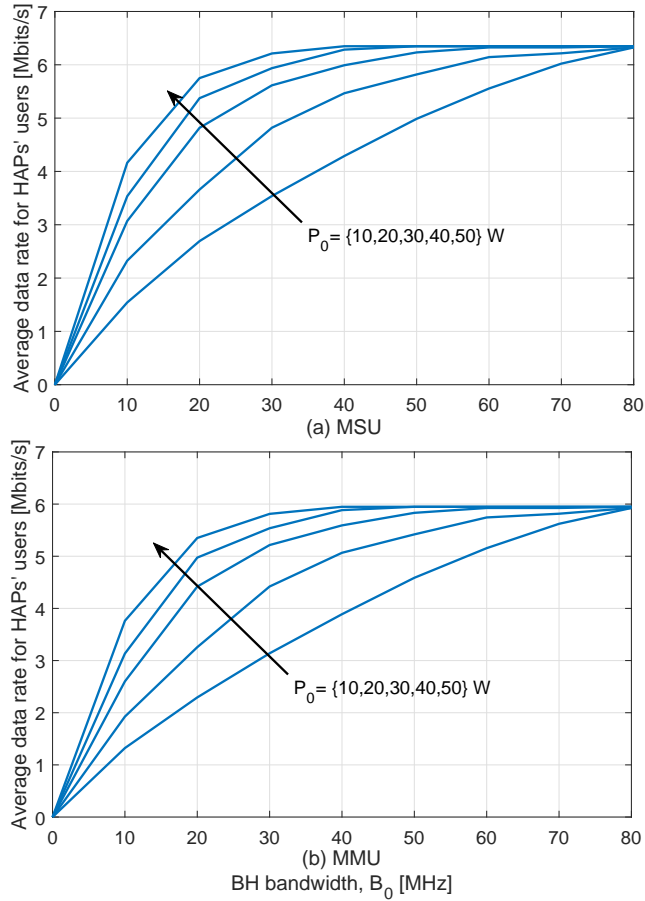}}
   \caption{Average data rate of HAPs' users versus BH bandwidth.}\label{fig2}
\end{figure}

\begin{figure}[h!]
  \centerline{\includegraphics[width=3.4in]{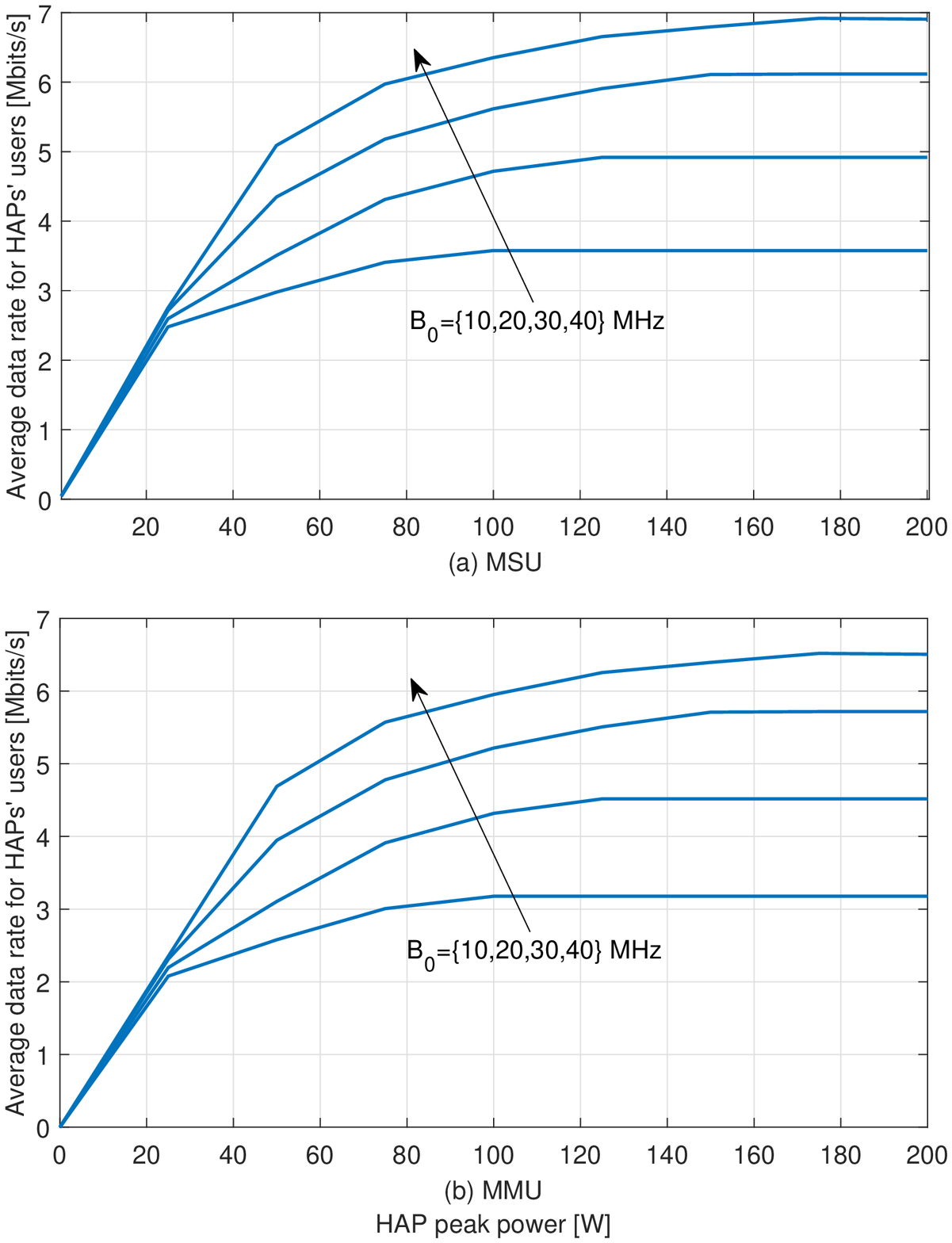}}
   \caption{Average data rate of HAPs' users versus HAPs' peak power.}\label{fig3}
\end{figure}

Fig.~\ref{fig2} and Fig.~\ref{fig3} illustrate the performance of the average data rate for users associated with HAPs using $U=400$.
For instance, Fig.~\ref{fig2} plots the effect of the BH bandwidth $B_0$ on the average data rate of HAPs' users for different BH transmit powers $P_0=\{10,20,30,40\}$ W with $\bar{P}_l=100$ W.
This figure shows that the average data rate of HAPs' users is improving with the increase of $B_0$ up to a certain cutoff value, that depends on the BH rate constraint as given in~\ref{rate_backhauling}.
This is due to the fact that starting from this point of $B_0$ the average data rate can not be enhanced further because
it depends on the value of $P^L_{lu,n}$ which is limited by $\bar{P}_L$ as given in~\eqref{powerm}. Also, Fig.~\ref{fig2} shows that the cutoff value depends on the BH transmit power $P_0$. As $P_0$ increases the corresponding $B_0$ value of the cutoff value decreases in order to respect constraint~\ref{rate_backhauling}. In this case, the bottleneck is the FH rate constraint which is limited by $\bar{P}_L$.

On the other hand, to illustrate the BH bottleneck, Fig.~\ref{fig3} plots the average data rate of HAPs' users versus HAPs' peak power $\bar{P}_L$ using $P_0=40$ W.
The figure shows that as $\bar{P}_l$ increases, the average data rate is increases up to a certain point. This can be explained by starting from this point of $\bar{P}_l$, the average data rate can not be improved because it depends also on
the BH data rate constraint $\bar{R}_{wl}$ as given in~\ref{rate_backhauling}. Furthermore, it can be noticed that the average data rate is enhanced as $B_0$ increases due to the reason, that increasing $B_0$ will also increase the value of $\bar{R}_{wl}$ and thus, increase the BH bottleneck.

\begin{figure}[h!]
  \centerline{\includegraphics[width=3.0in]{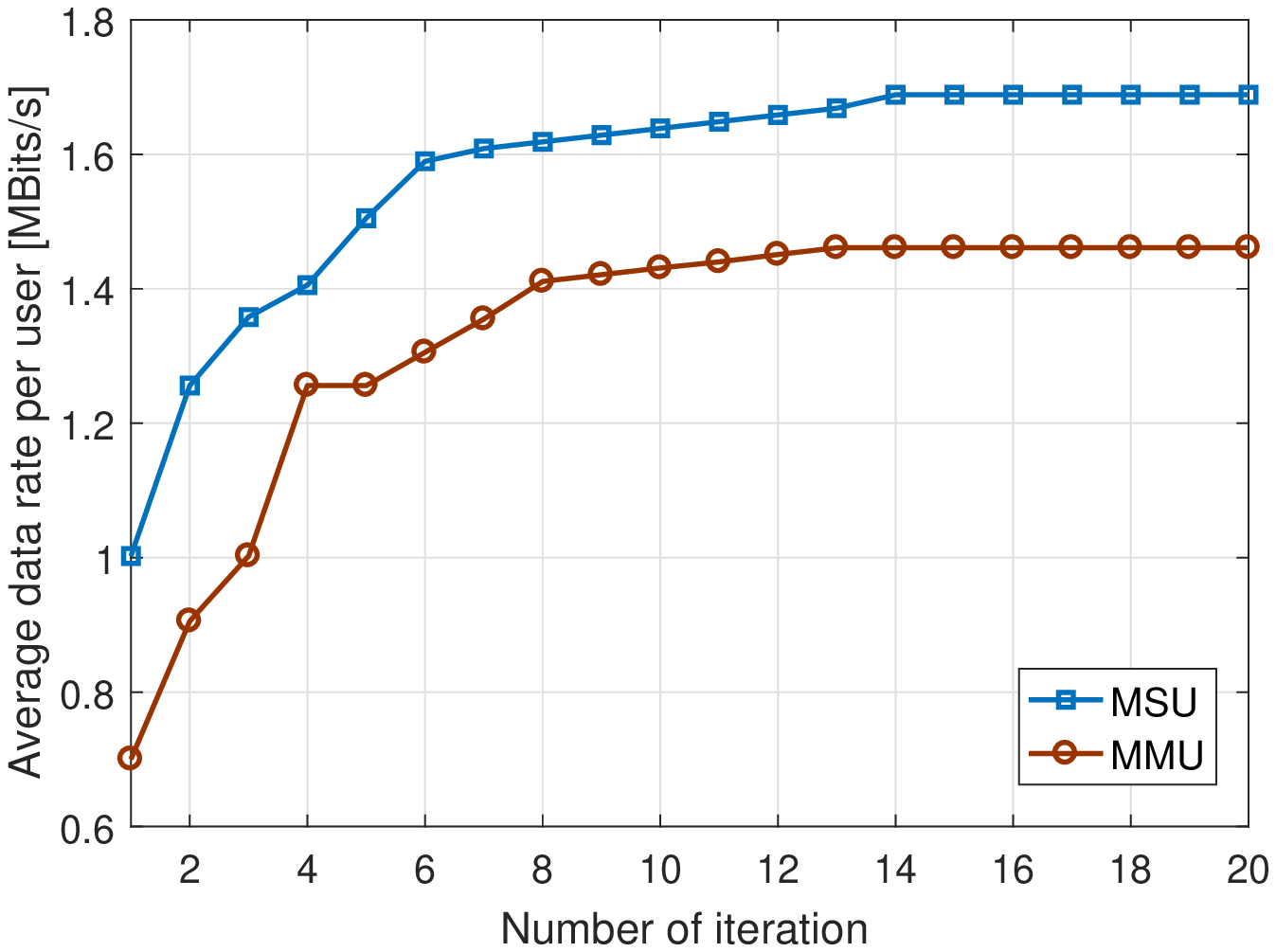}}
   \caption{Convergence of the proposed SR algorithm.}\label{convergence}
\end{figure}

Finally, Fig.~\ref{convergence} plots the convergence speed of the SR algorithm, defined by the number of
iterations needed to reach convergence, for both utilities, MSU and MMU. Note that one iteration in Fig.~\ref{convergence} corresponds to one iteration of the “while loop” given in Algorithm 2 line 4-13. It can be noticed that the algorithm is converge within around 8-13 iterations.

\section{Conclusion}\label{Conclusion}
This paper proposed an efficient optimization framework using TBSs, HAPs, and satellite station to provide connectivity to the ground users taking into consideration the BH limitation.
The objective was to maximize the users' throughput by optimizing the front-hauling and back-hauling associations, transmit powers of the base stations, and the HAPs' locations. We proposed an approximate and a low complexity solutions to optimally determine the decision variables.
The simulation results illustrated the behavior of our approach and their significant impacts on the users' throughput.
In our next challenging task, a free-space optical (FSO) communication link between GW and HAPs will be considered in order to mitigate the BH bottleneck limitation thus enhance the performance.
However, it will add more complexity to the problem by optimizing extra parameters such as the LoS angles.

\bibliographystyle{IEEEtran}
\bibliography{J_2019HAP}

\end{document}